\documentclass[conference]{IEEEtran}

\usepackage{amsmath}
\usepackage{graphicx}
\usepackage{subfig}
\usepackage{color}

\usepackage{algpseudocode}
\usepackage{algorithmicx}
\usepackage{framed}

\begin{document}
	
\title{Adaptive Work-Efficient Connected Components on the GPU}	

\author{\IEEEauthorblockN{Michael Sutton, Tal Ben-Nun, Amnon Barak}
\IEEEauthorblockA{Department of Computer Science\\
		Hebrew University of Jerusalem\\
		Jerusalem 91904, Israel\\
		Email: \{msutton,talbn\}@cs.huji.ac.il}
\and
\IEEEauthorblockN{Sreepathi Pai, Keshav Pingali}
\IEEEauthorblockA{Institute for Computational Engineering and Sciences\\
	    The University of Texas at Austin\\
		Austin, Texas, USA\\
		Email: sreepai@ices.utexas.edu, pingali@cs.utexas.edu}}
\maketitle

\begin{abstract}
This report presents an adaptive work-efficient approach for implementing the Connected Components algorithm on GPUs. The results show a considerable increase in performance (up to 6.8$\times$) over current state-of-the-art solutions.
\end{abstract}

\section{Introduction}

Finding the Connected Components (CC) of a graph is a fundamental algorithm, often used as a pre-processing step for other graph algorithms. 
Parallel CC algorithms exhibit fine-grain parallelization opportunities, as have been explored in the literature \cite{Awerbuch83, Shiloach82}. In particular, two parallel approaches are prominently used for CC, namely Label Propagation and ``Hook-Shortcut''. Soman et. al. \cite{soman} implemented a parallel GPU variant of Hook-Shortcut, showing that GPUs are beneficial for CC computation despite the inherent irregularity of the algorithm.

This report presents an adaptive variant of Hook-Shortcut CC, focusing on the two core operations of the algorithm and implementing them in a device-centric approach. By doing so, we improve the work-efficiency of the algorithm, increase its performance, and reduce CPU-GPU round-trips to minimum. We demonstrate that the algorithm is highly efficient on a wide variety of graph types.

\section{Existing Approaches}	

In graph theory, a connected component of an undirected graph $G = \{V, E\}$ is defined as a subset of vertices $\nu \subseteq V$, in which each two vertices are connected via an edge path. Computing the connected components $\{\nu_i\}_{i=1}^N$ of a graph, where $N$ is the number of components, is thus defined by assigning a unique component ID $C_i$ to each of the vertices $v\in \nu_i$. In practice, the component ID is determined by an index of some representative vertex in $\nu$.

Straightforward sequential solutions for CC are based on graph traversal, propagating the unique component ID, called label, via neighboring vertices until the entire component is labeled. Label propagation can be parallelized by initializing each vertex with a unique ID, disseminating vertex labels in parallel according to a global rule, such as minimal/maximal vertex index.

Shiloach and Vishkin \cite{Shiloach82} introduced a different parallel algorithm for CC, based on two operations called ``Hooking'' and ``Shortcutting''. Rather than propagating values through the graph, this variant transforms the input graph into trees, iteratively connecting those trees and reducing their depth. The algorithm converges when the graph is converted into a forest of depth-one trees, each representing a different component. Specifically, this variant defines two core procedures: (i) Hook, which, given an edge $\left(u,v\right)\in E$, attempts to connect the two trees that $u$ and $v$ belong to (if not already connected); and (ii) Shortcut, which replaces the parent of each tree vertex with its grand-parent, reducing the tree depth. The process alternates between the Hook and Shortcut operations until convergence.

Soman et al. \cite{soman} propose a variant of the Hook-Shortcut CC, where the Shortcut operation is replaced by Compression, which consecutively performs Shortcutting (called ``Jump'' operations in this variant) until each tree is reduced into a single-level ``star'' (depth-one tree). Hence, the algorithm iteratively alternates between Hook and Compress operations, reducing the number of iterations until convergence.

\begin{figure}[h]
	\begin{framed}
		\textbf{procedure} ConnectedComponents($G(V, E)$):
		\begin{algorithmic}[1]
			\State \textbf{let} $\pi : \pi(vertex) \gets vertex$
			\Repeat 
			\ForAll {$e$ in $E$} \textbf{in parallel} 
			\State $Hook (e, \pi)$ 
			\EndFor
			\Repeat 
			\ForAll {$v$ in $V$} \textbf{in parallel} 
			\State $Jump (v, \pi)$ 
			\EndFor
			\Until {$Jump$ performs no change in $\pi$}
			\Until {$Hook$ performs no change in $\pi$}
			\State \Return {$\pi$}
		\end{algorithmic} 
	\end{framed}
	\caption{Hook-Compress Connected Components}
	\label{alg:somancc-host}
\end{figure}

\begin{figure}[t]
		\begin{framed}
		\textbf{device procedure} Hook($edge(u, v), \pi$):
		\begin{algorithmic}[1] 
			\State $H \gets max \{  \pi(u), \pi(v)  \}$
			\State $L \gets min \{  \pi(u), \pi(v)  \}$
			\State $\pi(H) \gets L$
		\end{algorithmic}
		\bigbreak
		\textbf{device procedure} Jump($v, \pi$):
		\begin{algorithmic}[1] 
			\State $\pi(v) \gets \pi(\pi(v))$
		\end{algorithmic}
	\end{framed}
	\caption{Hook-Compress GPU Methods}
	\label{alg:somancc-dev}
\end{figure}

Fig. \ref{alg:somancc-host} depicts the Hook-Compress implementation for GPUs, proposed by Soman et al. \cite{soman}. The algorithm begins by defining a workspace $\pi$ for the component trees, initializing each vertex to point to itself (line 1). Then, the algorithm iteratively performs Hook rounds (lines 3-5) followed by one or more Jump calls (lines 6-10), flattening the trees by iteratively assigning $\pi(v)\leftarrow \pi(\pi(v))$ for each $v\in V$. 

The corresponding Hook-Compress GPU kernels are presented in Fig. \ref{alg:somancc-dev}. As seen in the figure, although parallel Hook calls may perform concurrent writes to the same memory locations (Hook procedure, line 3), its implementation is atomic-free (i.e., does not use atomic operations). This is supported by the overall algorithm logic, which assures every Hook operation will eventually succeed before converging. The Hook procedure also uses a global high-to-low hooking rule for avoiding possible tree cycles. \newline

\section{Work-Efficient Connected Components}

This section proposes a work-efficient approach for implementing Hook-Compress CC on the GPU. This approach redefines the two core operations of the algorithm, Hook and Compress, moving the control-flow to the GPU. We show that the variant is scalable, minimizes CPU-GPU communication, adaptive with respect to graph properties (e.g., average degree), and converges faster than existing approaches. Below, we describe the algorithm and elaborate on its work-efficiency and adaptability properties.

\subsection{Work-Efficiency}

The proposed CC algorithm is composed of two GPU kernels: Multi-Jump and Atomic-Hook.

\algblockdefx[]{Lock}{EndLock}
[1] {\textbf{lock} #1}
[0] {\textbf{end lock}}

\begin{figure}[t]
	\begin{framed}
		\textbf{device procedure} AtomicHook($edge(u, v), \pi$):
		\begin{algorithmic}[1] 
			\While {$\pi(u) \neq \pi(v)$}
			\State $H \gets max \{  \pi(u), \pi(v)  \}$
			\State $L \gets min \{  \pi(u), \pi(v)  \}$
			\Lock {$\pi(H)$} \Comment(atomic CAS)
			\If {$\pi(H) = H:$}
			\State $\pi(H) \gets L$  
			\State \Return
			\Else
			\State $u \gets \pi(H)$
			\State $v \gets L$
			\EndIf
			\EndLock
			\EndWhile
		\end{algorithmic}
		\bigbreak
		\textbf{device procedure} MultiJump($v, \pi$):
		\begin{algorithmic}[1] 
			\While {$\pi(\pi(v)) \neq \pi(v)$}
			\State $\pi(v) \gets \pi(\pi(v))$
			\EndWhile
		\end{algorithmic}
	\end{framed}
	\caption{Atomic-Hook and Multi-Jump Kernels}
	\label{alg:ahcc}
\end{figure}

In \textbf{Multi-Jump}, instead of performing multiple single-level Jump operations, we fuse an entire \textit{Compress} phase into a single kernel. This kernel, shown in the bottom portion of Fig. \ref{alg:ahcc}, replaces lines 6-10 in Fig. 1 with a single parallel invocation. 

We note two important optimizations in our Multi-Jump procedure: (i) the procedure constantly writes an updated value to $\pi(v)$ even though one final write would suffice; and 
(ii), the parallel kernel is scheduled with partial order, starting from the top-most vertices in the component trees (lower indices) and going down to the leaves (higher indices).
The two optimizations decrease the number of iterations each device-thread must perform, hence reducing thread-divergence introduced by the irregular while loop.  

The second kernel is an \textbf{Atomic-Hook} procedure. This kernel uses atomic Compare-And-Swap (CAS) operations to traverse the component trees until a hook operation can be verified to succeed, without overriding other concurrent hook operations. This eliminates the need for the CPU-side convergence loop in lines 2-11 of Fig. \ref{alg:somancc-host}.

The Atomic-Hook procedure, depicted in the top portion of Fig. \ref{alg:ahcc}, begins by finding the parents of both $(u, v)$ and determining their high-low order for global consensus (lines 1-3). Then, with an atomic CAS operation, $\pi(H)$ is locked and checked; if $H$ is the root of a tree, then it can be used for hooking the entire tree without overriding any other pointer, hence the procedure assigns $L$ to be its parent and exits successfully (lines 5-7). Otherwise, the procedure recursively considers $(\pi(H), L)$ to be the new $(u, v)$ edge to connect (lines 9-10) at the next iteration. The recursive logic continues until successfully acquiring a root (line 5), or if discovering the two component trees are already connected (line 1).   

\subsection{Adaptability}

\label{sec:adapt}

Using the above two kernels, one could possibly solve CC with only two kernel calls: a single Atomic-Hook (guaranteeing component tree connectivity), followed by a Multi-Jump operation for compressing all component trees into ``stars''.

Performing a single Atomic-Hook may be more efficient than several non-atomic Hooks, but it eliminates the ability to perform constant Compress operations between each Hook round, which may degrade the overall performance. 
For enabling intermediate tree compressions we introduce segmentation, i.e. splitting the input graph into distinct edge-list segments and performing Compress operations between atomic segment Hooks. 
Below we describe a heuristic method for selecting the optimal number of edge-list segments to split into.

We highlight two observations, both related to the ratio between the number of edges $|E|$, and the number of vertices $|V|$ (i.e. the average vertex degree) of the input graph. 

First, atomic operations within the Atomic-Hook procedure are performed on the component trees workspace, which has $|V|$ memory size. As a result, the true parallelism of atomic operations is bounded by the number of vertices, creating a growing atomic contention as average degree increases. 

Second, unlike the Hook step, which has $O(|E|)$ complexity, the Compress step is performed over vertices and hence has $O(|V|)$ complexity. As the number of edges grows compared to the number of vertices, the cost of Compress operations becomes a minor percentage of the overall run-time cost, and performing more intermediate compressions may be beneficial. 

Based on these observations, we suggest a heuristic method to determine the best segmentation, is to split the graph into $2|E| / |V|$ segments (the average degree), effectively performing Atomic-Hook operations over edge-list segments of size $\approx |V|$ each followed by a full $O(|V|)$ Compress.

This way, Hook operations are always performed over segments proportional to the $|V|$ memory workspace, reducing atomic contention. Each such segment Hook, is followed by a Compress operation, flattening all component trees and minimizing atomic operations for the next segment Hook.
As average degree increases, more Compress operations will be performed, but their $O(|V|)$ cost becomes minor compared to the dominating $O(|E|)$ complexity of the overall algorithm.

\begin{figure}[h]
	\begin{framed}
		\textbf{procedure} AdaptiveConnectedComponents($G(V, E)$):
		\begin{algorithmic}[1]
			\State \textbf{let} $\pi : \pi(vertex) \gets vertex$
			\State \textbf{let} $s \gets 2|E| / |V|$ \Comment (average degree)
			\State \textbf{let} $\{E_i\}_{i=1}^s$ be $s$ distinct subsets of $E$
			\For{$i\gets 1, s$}
			\ForAll {$e$ in $E_i$} \textbf{in parallel} 
			\State $AtomicHook (e, \pi)$ 
			\EndFor
			\ForAll {$v$ in $V$} \textbf{in parallel} 
			\State $MultiJump (v, \pi)$ 
			\EndFor
			\EndFor
			\State \Return {$\pi$}
		\end{algorithmic} 
	\end{framed}
	\caption{Adaptive Hook-Compress}
	\label{alg:adaptcc-host}
\end{figure}

Fig. \ref{alg:adaptcc-host} shows the final host-side pseudocode for Adaptive Connected Components.

\section{Performance Evaluation}

\begin{table}[h]
	\caption{Graph Properties}
	\label{tbl:graphtypes}
	\centering
	\scriptsize
	\begin{tabular}{|l|c|c|c|c|c|}
		\hline
		\bf    Name & \bf Nodes & \bf Edges & \bf Avg.   & \bf Max & \bf Size \\
					&           &           & \bf Degree & \bf Degree & \bf  (GB)\\\hline 
		\multicolumn{6}{|c|}{\bf Road Maps}\\\hline
		usa-osm \cite{dimacs9}         & 24M & 58M & 2.41 & 9 & 0.62 \\\hline
		euro-osm-karls \cite{karlsruhe}   & 174M & 348M & 2.00 & 15 & 3.90 \\\hline
		\multicolumn{6}{|c|}{\bf Social Networks}\\\hline
		soc-live-journal \cite{soclj} & 5M & 69M & 14.23 & 20,293 & 0.56 \\\hline
		\multicolumn{6}{|c|}{\bf Synthetic Graphs}\\\hline
		kron-logn21 \cite{dimacs10} & 2M & 182M  & 86.82 & 213,904 & 1.40 \\\hline
	\end{tabular}
\end{table}

Our experimental setup consists of an NVIDIA Tesla M60 (Maxwell architecture) GPU, containing 16 multiprocessors with 128 cores each. We evaluated four graph instances of different types, as listed in Table \ref{tbl:graphtypes}. 

\begin{figure}[t]
	\centering
	\includegraphics[trim={5.75cm 4.25cm -1cm 3.25cm},clip,height=1.7in]{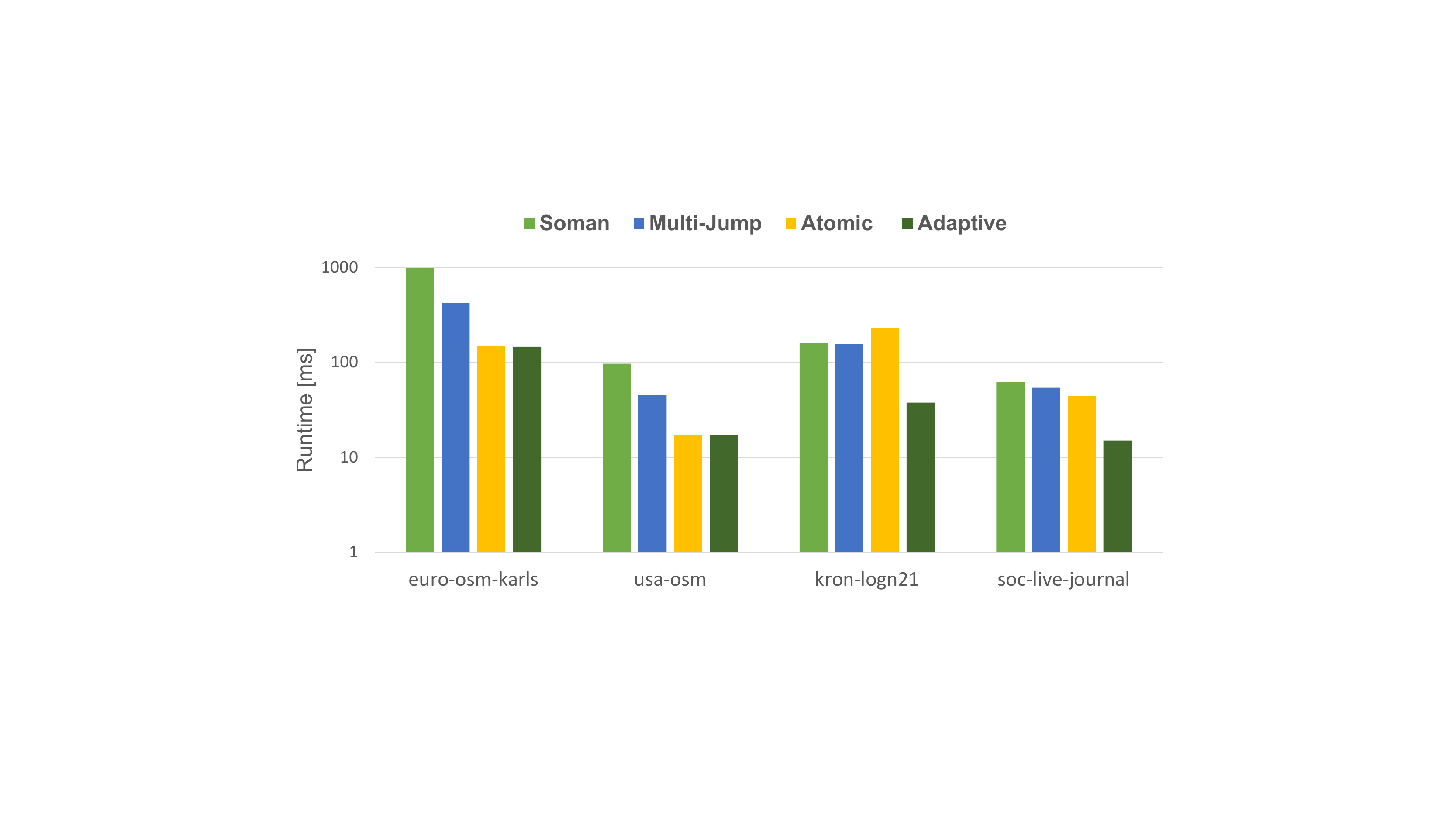}
	\caption{Adaptive Connected Components Performance (lower is better)}
	\label{fig:overall-throughput}
\end{figure} 

Fig. \ref{fig:overall-throughput} compares the baseline Soman \cite{soman} code with our approach, gradually adding Multi-Jump, Atomic-Hook, and Adaptability. The figure shows that the adaptive version consistently outperforms the other versions. Observe that Multi-Jump improves performance in all graph types, whereas Atomic-Hook exhibits improvement when adapting the number of segments based on the given input graph. For the best segmentation, we achieve consistent speedups over the current state-of-the-art \cite{soman} ranging from 4.15$\times$ to 6.76$\times$.  

Also note that for graphs like \textit{kron-logn21} and \textit{soc-live-journal} \cite{dimacs10, soclj} which have a relatively high average degree, the effect of the Multi-Jump optimization is minor (1.02$\times$ and 1.14$\times$ respectively) because it only affects Compress operations, which have a small $O(|V|)$ complexity in these cases.

\begin{figure}[h]
	\centering
	\includegraphics[trim={6.25cm 4.75cm 4cm 4.25cm},clip,height=1.7in]{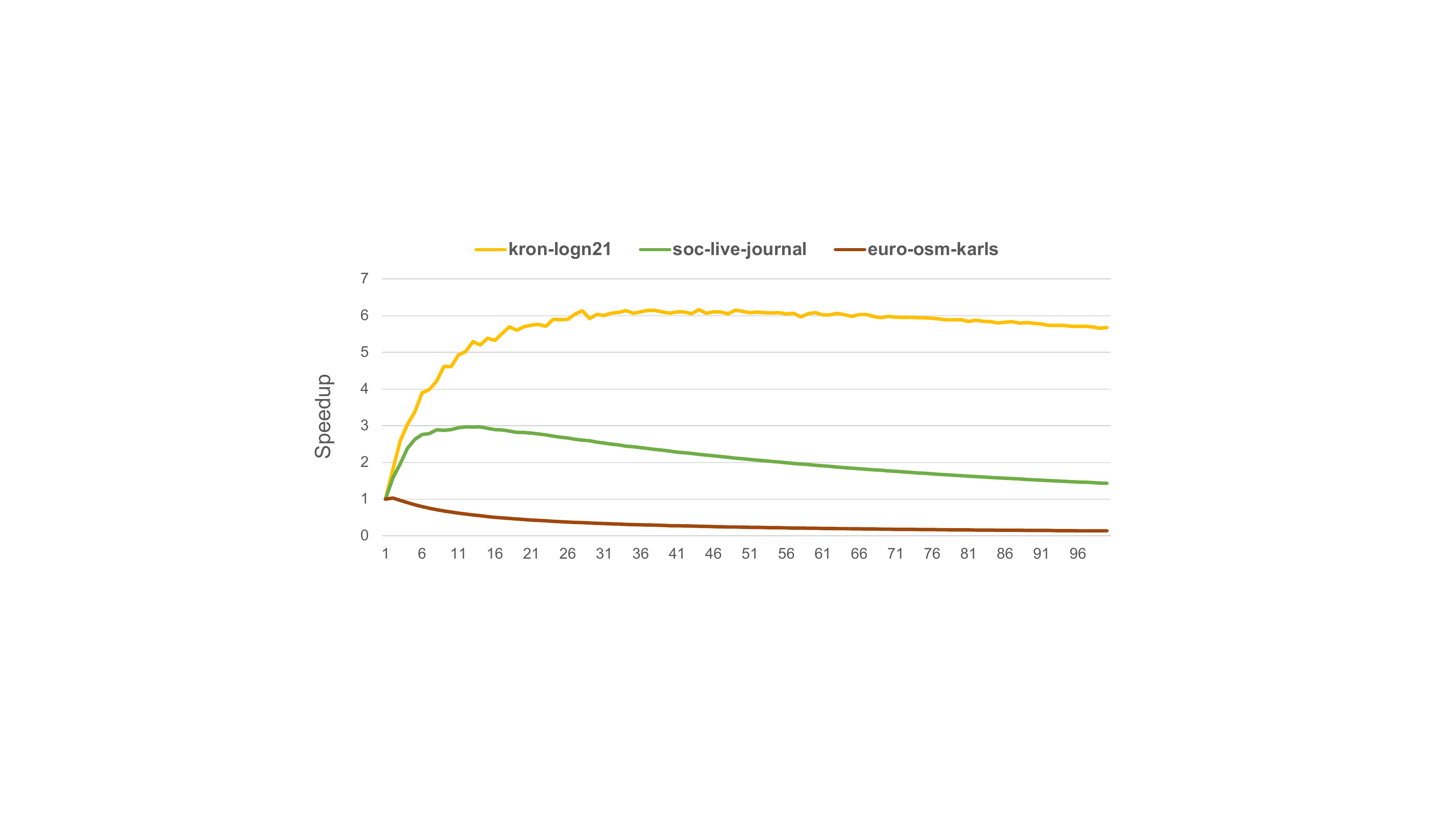}
	\caption{Segmentation Speedup}
	\label{fig:seg-speedup}
\end{figure}

Fig. \ref{fig:seg-speedup} shows the segmentation speedup over the Atomic-Hook baseline (single segment) for various graphs on the average degree spectrum. For all tested graphs, the figure shows that the speedup is maximal when segmenting into $2|E| / |V|$ segments. These results coincide with the explanation behind the adaptive heuristic, as described in Section \ref{sec:adapt}.

\section*{Acknowledgment}
This research was supported by the German Research Foundation (DFG)
Priority Program 1648 ``Software for exascale Computing'' (SPP-EXA),
research project FFMK; NSF grants 1218568, 1337281,
1406355, and 1618425; by DARPA BRASS contract
750-16-2-0004; and an equipment grant from NVIDIA.

\bibliographystyle{IEEEtran}
\bibliography{IEEEabrv,references}

\end{document}